\documentstyle[preprint,aps,psfig]{revtex}

\begin{document}

\draft
\preprint{
\vbox{
\hbox{TUM/T39-95-9}  
\hbox{December 1995}
}}

\title{Next-To-Leading Order Analysis of Polarized and Unpolarized
       Structure Functions\footnote[1]{Work supported in part 
       by BMBF and by DFG grant We 655/11-2}}

\author{T.Weigl$^a$ and W.Melnitchouk$^{a,b}$}

\address{$^a$   Physik Department,
                Technische Universit\"{a}t M\"unchen,
                D-85747 Garching, Germany.}

\address{$^b$   Department of Physics,
                University of Maryland,
                College Park, MD  20742, USA.}

\maketitle

\begin{abstract}
We present numerical solutions of the $Q^2$ evolution
equations at next-to-leading order (NLO) for unpolarized and
polarized parton distributions, in both the flavor non-singlet
and singlet channels.
The numerical method is based on a contour integration algorithm
in complex-moment space, and, unlike most standard fitting techniques,
is not restricted to the use of analytic moments of distributions.
We also compute analytic continuations of the recently
calculated singlet anomalous dimensions, needed for evolving
polarized quark and gluon distributions at NLO.
Using these we study the NLO effects on the $x$ dependence of the
spin-dependent structure functions $g_1^p$ and $g_1^n$.
\end{abstract}

\pacs{PACS numbers: 13.60.Hb, 13.88.+e, 14.20.Dh        \\ \\
      Submitted to {\em Nuclear Physics B}.}

\section{Introduction}

One of the original testing grounds for QCD has been the study of 
scaling violations in structure functions in nucleon deep-inelastic
scattering (DIS).
Over the last 25 years a wealth of information has been accumulated
on the unpolarized structure functions, $F_{1,2}$, covering some
three orders of magnitude of the momentum transfer squared, $Q^2$,
and probing the small Bjorken-$x$ region down to
$x = Q^2/2M\nu \sim 10^{-4}$, where $\nu$ is the energy transfer
to the nucleon at rest, and $M$ the nucleon mass.
The data at large $Q^2$ ($Q^2 \agt 5$ GeV$^2$) show no deviation from 
the perturbative QCD (pQCD) predictions obtained from solutions of 
the Dokshitzer-Gribov-Lipatov-Altarelli-Parisi (DGLAP) evolution 
equations \cite{DGL,AP77}.
More recently, accurate data \cite{SMCP,E142} have also been 
accumulating on the spin-dependent structure functions, $g_{1,2}$,  
the $Q^2$ dependence of which will be necessary to confront the 
pQCD expectations.

The need for an accurate description of the $Q^2$ dependence of
both unpolarized and polarized nucleon structure functions is
severalfold.
Firstly, in modeling hard processes other than DIS, the ingredients
needed to calculate cross sections and event rates are parton
distribution functions at a certain scale $Q^2$.
Because the energies involved in purely hadronic reactions are
often considerably larger than in DIS experiments, it is imperative
that evolution of the distributions (which are extracted primarily
from DIS data) to the higher $Q^2$ be as reliable as possible.

At the other end of the energy spectrum, describing the $Q^2$
behavior of structure functions at low $Q^2$ is no less important.
While the $Q^2$ dependence of parton distributions is accessible 
to pQCD, their $x$ dependence is of course not directly calculable.
Often one tries to connect the non-perturbative content of parton 
distributions to valence quark models of QCD \cite{JR}, which are 
presumed to be good approximations at $Q^2 \alt 1 \mbox{GeV}^2$ 
(for a recent summary see Ref.\cite{CQ} and references therein).
To compare with data one usually evolves the leading twist component
to the higher $Q^2$ of the DIS experiments.
However, because this procedure involves evolution through a region
where the strong running coupling constant, $\alpha_s$, is no longer 
small, one should at least take next-to-leading order (NLO) 
corrections into account before seriously testing the reliability of 
these calculations.

In the spin-dependent sector, while the kinematics in experiments
involving polarization are more limited than in unpolarized processes
(average $Q^2$ values in recent $g_1$ measurements only vary between
2 and 10 GeV$^2$), knowledge of the $Q^2$ dependence of polarized
structure functions is essential in connection with questions
concerning the spin content of the nucleon.
Aside from accurately correcting for the scale dependence of first
moments of polarized distributions \cite{Q2MOM}, such as in the
Bjorken sum rule, one also needs to understand the effects of
$Q^2$ evolution upon the {\em shape} of the proton, and in particular
neutron, $g_1$ structure functions themselves \cite{XDEP}.
While the leading order pQCD corrections to $g_1$ have been known 
for some time, the NLO effects have until now only been 
calculated for non-singlet (NS) distributions.
Results recently published by Mertig and van Neerven \cite{MN95}
for the anomalous dimensions of singlet operators at NLO have 
finally made it possible to compute the NLO corrections to the 
polarized singlet quark and gluon distributions as well.

In this paper we analyze the effects of NLO corrections on polarized
and unpolarized parton distributions, in both the flavor singlet and
non-singlet channels, as a function of $x$ and $Q^2$.
In particular, we concentrate on the $Q^2$ evolution of the nucleon
$g_1$ structure function, and its effect on the $x$ dependence at
small and intermediate $x$.
To study the effects numerically, we develop an accurate $Q^2$
evolution algorithm based on an inverse Mellin transform method.

Often one encounters the need to evolve distributions whose 
moments are not able to be expressed in simple analytic forms such  
as those used in standard data fitting in Refs.\cite{MRS,CTEQ,GR90}.
For example, most studies of structure functions in low-energy
models of the nucleon, including recent attempts \cite{KRE} to   
describe the nucleon's non-perturbative sea within the pion-cloud 
model \cite{PIONS}, can only be expressed numerically, 
with the output needing to be parametrized before
the standard evolution and fitting techniques can be applied.  
To facilitate evolution of both 
parametrized and calculated distributions, 
our method therefore provides for the possibility of evolving 
input distributions given only at a limited number
of points in Bjorken $x$.

The paper is organized as follows.
In Section II we review the essential features of the DGLAP evolution
equations at ${\cal O}(\alpha_s^2)$, summarizing for convenience 
the relevant formulas 
for evolution of parton distributions and structure functions.
More comprehensive discussions can be found for example in
Refs.\cite{FP82,AE78,A82}.
A detailed description of the numerical method employed to
perform the evolution is given in Section III.
The numerical results, for both unpolarized and polarized DIS, 
are presented in Section IV, and conclusions are drawn in Section V.
Finally, in the Appendices we collect all the relevant formulas
for the LO and NLO anomalous dimensions, together with details 
about their extrapolation into complex moment space.

\section{Evolution Equations at Next-To-Leading Order}

In this Section we review the main features pertinent to the DGLAP
evolution equations, including ${\cal O}(\alpha_s^2)$ corrections.
We begin with the usual forward Compton amplitude, $T_{\mu\nu}$, 
for the scattering of a virtual photon from a nucleon,
\begin{equation}
T_{\mu\nu}
= \frac{i}{2M} \int d^4z\ e^{iq\cdot z}
\langle p, S \left|
T \left( J_{\mu}(z) J_{\nu}(0) \right)
\right| p, S \rangle,
\end{equation}
where $p$ and $q$ are the nucleon and virtual photon four-momenta,
and $S$ is the nucleon spin vector.
The fact that in the Bjorken limit ($p \cdot q, -q^2 \rightarrow
\infty$) the amplitude $T_{\mu\nu}$ is light-cone dominated
allows for an operator product expansion (OPE) analysis of the
time-ordered product $T \left( J_{\mu}(z) J_{\nu}(0) \right)$,
through which the long and short distance physics are formally 
separated.
The singular bilocal operator $T \left( J_{\mu}(z) J_{\nu}(0) \right)$
is expanded in a basis of regular local operators, with the light-cone
singularity contained in the Wilson expansion coefficient functions
\cite{Wi69,Bu80,Ro90}.

The forward Compton amplitude $T_{\mu\nu}$ is related to the nucleon
DIS hadronic tensor $W_{\mu\nu}$ through the optical theorem:
\begin{equation}
W_{\mu\nu} = \frac{1}{\pi} \Im m T_{\mu\nu}.
\end{equation}
The hadronic tensor is usually written in terms of the spin-independent
$F_1$ and $F_2$, and spin-dependent $g_1$ and $g_2$ structure functions as:
\begin{eqnarray}
\label{Htensor}
W_{\mu\nu}
&=& \frac{F_1(x,Q^2)}{M}
    \left( \frac{q_{\mu}q_{\nu}}{q^2} - g_{\mu\nu} \right)
 + \frac{F_2(x,Q^2)}{M p\cdot q}
\left( p_{\mu} - \frac{p\cdot q}{q^2} q_{\mu} \right)
\left( p_{\nu} - \frac{p\cdot q}{q^2} q_{\nu} \right)
\nonumber\\
&+& i \epsilon_{\mu\nu\alpha\beta} q^{\alpha}
    \left( {S^{\beta} \over p \cdot q}
           \left( g_1(x,Q^2) + g_2(x,Q^2) \right)
         - p^{\beta} {S \cdot q \over (p \cdot q)^2} g_2(x,Q^2)
    \right),
\end{eqnarray}
where $x = Q^2/2 p \cdot q$ and $Q^2 = -q^2$.
The structure functions are related to parton distribution functions
(light-cone momentum distributions) by convoluting the latter with 
the Wilson coefficient functions \cite{Bu80,CH72,Mu87}, describing
the hard (perturbatively calculable) photon--parton interaction.  
For the proton's $F_2$ structure function, for instance, one has:
\begin{eqnarray}
\label{F2con}
F_2^p(x,Q^2)
&=& {1 \over 6}
    \int_x^1\ {dy \over y}
    \left( C_2^{NS}\left(x,Q^2\right)\
           x q^{NS}\left( {x\over y},Q^2 \right)
    \right.                                     \nonumber\\
& & \left.
        +\ {5 \over 3} C_2^q \left(x,Q^2\right)\
           x \Sigma\left( {x \over y},Q^2\right)
        +\ {5 \over 3} C_2^G \left(x,Q^2\right)\
           x G\left( {x \over y},Q^2\right)
    \right),
\end{eqnarray}
where $q^{NS}(x,Q^2) = (u + \bar u - d - \bar d - s - \bar s)(x,Q^2)$
is the flavor non-singlet combination (for three flavors), while the
singlet component is $\Sigma(x,Q^2) = \sum_q (q + \bar q)(x,Q^2)$,
and $G(x,Q^2)$ is the gluon distribution.

Similarly for polarized DIS, one has for the $g_1$ structure
function of the proton:
\begin{eqnarray}
\label{g1con}
g_1^p(x,Q^2)
&=& 
    \int_x^1\ {dy \over y}
    \left( \Delta C^{NS}(x,Q^2)
           \left[ {1\over 12} \Delta q_3\left( {x\over y},Q^2 \right)
                + {1\over 36}
                  \Delta q_8\left( {x\over y},Q^2 \right)
           \right]
    \right.                     \nonumber\\
& & + \left.
           {1 \over 9}
           \Delta C^q\left(x,Q^2\right)\
           \Delta \Sigma\left( {x\over y},Q^2 \right)
         + {1\over 9}
           \Delta C^G\left(x,Q^2\right)\
           \Delta G\left( {x\over y},Q^2 \right)
    \right),
\end{eqnarray}
where the polarized singlet distribution is
$\Delta \Sigma(x,Q^2)
= \sum_q (\Delta q + \Delta \bar q)(x,Q^2)$,
$\Delta G(x,Q^2)$ is the polarized gluon distribution, 
and the non-singlet combinations $\Delta q_3$ and $\Delta q_8$
are (for three flavors) given by:
\begin{mathletters}
\begin{eqnarray}
\Delta q_3(x,Q^2)
&=& \left( \Delta u + \Delta\overline{u}
         - \Delta d -\Delta\overline{d} \right)(x,Q^2),\\
\Delta q_8(x,Q^2)
&=& \left( \Delta u + \Delta\overline{u}
         + \Delta d  +\Delta\overline{d}
         - 2\left(\Delta s + \Delta\overline{s} \right)
    \right)(x,Q^2).
\end{eqnarray}
\end{mathletters}%
Note that the polarized quark singlet coefficient function
$\Delta C^q(x,Q^2) = \delta(1-x) + {\cal O}(\alpha_s)$,
while the polarized gluon enters only at NLO,
$\Delta C^G(x,Q^2) = {\cal O}(\alpha_s)$.
We shall return to this point in Section \ref{polRES}.

\subsection{Evolution of Parton Distributions}
\label{evpdf}

The $Q^2$ evolution of parton distribution functions is most
naturally expressed in terms of their moments, by solving the
renormalization group equations.
Defining the $n$th Mellin moment of a parton distribution
$q(x,Q^2)$ by:
\begin{equation}
\label{Mellin}
{\cal Q}_n(Q^2) = \int_0^1 dx\ x^{n-1} q(x,Q^2),
\end{equation}
one obtains for the NS moments:
\begin{equation}
\label{evoNS}
{\cal Q}_n^{NS}(Q^2)
= \left[1+\frac{\alpha_s(Q^2)-\alpha_s(\mu^2)}{4\pi}
  \left(\frac{\gamma_{NS}^{(1),n}}{2\beta_0} -
  \frac{\beta_1\gamma_{NS}^{(0),n}}{2\beta_0^2} \right)\right]
  \left(\frac{\alpha_s(Q^2)}{\alpha_s(\mu^2)}
  \right)^{\gamma_{NS}^{(0),n} / 2\beta_0}
{\cal Q}_n^{NS}(\mu^2),
\end{equation}
where $\gamma_{NS}^{(0,1),n}$ are the anomalous dimensions
at leading and next-to-leading orders, and $\mu^2$ is some
reference scale.
The constants $\beta_0$, $\beta_1$ are the expansion coefficients
of the Gell-Mann--Low function $\beta(Q^2)$ up to next-to-leading
order:
\begin{equation}
\beta_0 = 11-\frac23N_f,\qquad \beta_1 = 102 -\frac{38}{3}N_f,
\end{equation}
for $N_f$ active quark flavors.
Note that Eq.(\ref{evoNS}) is expressed within the
$\overline{\mbox{MS}}$ factorization scheme --- for the connection
to other factorization schemes see Appendix \ref{fact}.

Because of mixing between local operators of the OPE which are 
singlets under SU$(N_f)$ transformations, one has coupled evolution
equations for the singlet parton densities:
\begin{eqnarray}
\label{evoS}
\left(\begin{array}{c}
\Sigma_n(Q^2) \\ {\cal G}_n(Q^2) \end{array} \right)
&=& \left\{\left(\frac{\alpha_s(Q^2)}{\alpha_s(\mu^2)}
           \right)^{\lambda_-^n/2 \beta_0}
\left[ {\bf P}_-^n -\frac{1}{2\beta_0}\frac{\alpha_s(\mu^2) -
\alpha_s(Q^2)}{4\pi} {\bf P}_-^n {\bf R}^n {\bf P}_-^n \right.
\right.\nonumber\\
& & \left.
  - \left( \frac{\alpha_s(\mu^2)}{4\pi} - \frac{\alpha_s(Q^2)}{4\pi}
    \left(\frac{\alpha_s(Q^2)}{\alpha_s(\mu^2)}
    \right)^{(\lambda_+^n - \lambda_-^n) / 2\beta_0} \right)
\frac{{\bf P}_-^n {\bf R}^n {\bf P}_+^n}
     {2\beta_0+\lambda_+^n - \lambda_-^n} \right]
\nonumber\\
& & \left. + (+\longleftrightarrow -) \right\}
\left(\begin{array}{c}
\Sigma_n(\mu^2) \\ {\cal G}_n(\mu^2) \end{array}
\right),
\end{eqnarray}
where ${\bf R}^n,{\bf P}_{\pm}^n$ and $\lambda_{\pm}$ are
given by:
\begin{mathletters}
\begin{eqnarray}
{\bf R}^n
& = & \mbox{\boldmath$\gamma$}^{(1),n} -\frac{\beta_1}{\beta_0}
 \mbox{\boldmath$\gamma$}^{(0),n}, \\
{\bf P}_{\pm}^n
& = & \pm \frac{\mbox{\boldmath$\gamma$}^{(0),n}
  -\lambda_{\mp}^n}{\lambda_+^n - \lambda_-^n}, \\
\lambda_{\pm}^n
& = & \frac12 \left( {\gamma_{qq}^{(0),n} + \gamma_{GG}^{(0),n}
\pm \sqrt{(\gamma_{GG}^{(0),n} - \gamma_{qq}^{(0),n})^2 +
4\gamma_{qG}^{(0),n} \gamma_{Gq}^{(0),n}} }\right), \\
\mbox{\boldmath$\gamma$}^{(k),n} & = & \left(
        \begin{array}{cc} \gamma_{qq}^{(k),n} & \gamma_{qG}^{(k),n}\\
              {}              &         {}            \\
        \gamma_{Gq}^{(k),n} & \gamma_{GG}^{(k),n} \\
        \end{array}  \right), \ \ k=0,1 .
\end{eqnarray}
\end{mathletters}%

For polarized parton distributions one has identical evolution
equations to those in Eqs.(\ref{evoNS}) and (\ref{evoS}), with
${\cal Q}_n, \Sigma_n, {\cal G}_n$ replaced by their spin-dependent
counterparts $\Delta {\cal Q}_n, \Delta \Sigma_n, \Delta {\cal G}_n$.
The differences in the evolution arise from the anomalous dimensions,
which are in general different for polarized and unpolarized processes.
The explicit expressions for these at LO and NLO are given in
Appendix \ref{APP}.

The anomalous dimensions are also related to the Mellin
moments of the splitting functions $P_{ij}(z)$ ($i,j = q,G$):
\begin{eqnarray}
\gamma_{ij}^n
&\sim& \int_0^1 dz\ z^{n-1}\ P_{ij}(z),
\end{eqnarray}
where $P_{ij}(z)$ gives the light-cone probability of finding a
parton $i$ inside parton $j$, carrying a fraction $z$ of the parent
parton's light-cone momentum.
If the splitting functions (or the anomalous dimensions) are known,
then once the input parton distribution is fixed at a scale $\mu^2$, 
the $Q^2$ dependence of its moments can be calculated via
Eqs.(\ref{evoNS}) and (\ref{evoS}).
The $x$-dependent distribution at the new scale $Q^2$ is then
obtained by inverting the Mellin transform in Eq.(\ref{Mellin}).

\subsection{Evolution of Structure Functions}
\label{evsf}

{}From the evolved parton distributions one can obtain the
measured structure functions in two ways.
Firstly, one can evolve the parton distributions,
then convolute the result with the corresponding Wilson
coefficient functions, as in Eqs.(\ref{F2con}) and (\ref{g1con}).
In practice, however, to accurately determine the convolution 
integral requires interpolating the evolved distribution,
thereby introducing an additional source of numerical error.
Alternatively, the evolved moments can be multiplied by the
corresponding moments of the Wilson coefficient functions, before 
reconstructing the structure function via the inverse Mellin transform.
In this work we adopt this second method to evaluate the structure
function evolution.

The explicit relations between the moments of the non-singlet and 
singlet parts of the $F_2$ structure function and the corresponding 
moments of the parton distributions are:
\begin{eqnarray}
{\cal F}_{2,n}^{NS}(Q^2)
&=& C_{2,n}^q(Q^2)\ {\cal Q}_n^{NS}(Q^2), \\
{\cal F}_{2,n}^S(Q^2)
&=& C_{2,n}^q(Q^2)\ \Sigma_n(Q^2)\
 +\ C_{2,n}^G(Q^2)\ {\cal G}_n(Q^2).
\end{eqnarray}
For the $F_3$ structure function, in the case of
neutrino--nucleon scattering, one has:
\begin{eqnarray}
{\cal F}_{3,n}(Q^2)
&=& C_{3,n}^q(Q^2)\ {\cal Q}_n^V(Q^2),
\end{eqnarray}
where ${\cal Q}_n^V(Q^2)$ represents the $n$th moment of the
total valence distribution.
Up to NLO in the $\overline{\mbox{MS}}$ scheme the Wilson coefficients
are given by:
\begin{eqnarray}
C_{k,n}^{NS}(Q^2)
&=& C_{k,n}^q(Q^2)
 = C_{k,n}^{(0),q}
 + \frac{\alpha_s(Q^2)}{4\pi} C_{k,n}^{(1),q},  \\
C_{k,n}^G(Q^2)
&=& C_{k,n}^{(0),G}
 + \frac{\alpha_s(Q^2)}{4\pi} C_{k,n}^{(1),G},\ \ \ \ k=2, 3,
\end{eqnarray}
where
\begin{mathletters}
\begin{eqnarray}
C_{2,n}^{(0),q}
&=& 1,  \\
C_{2,n}^{(1),q}
&=& C_F \left[ 2S_1^2(n) - 2S_2(n) +3S_1(n)-\frac{2S_1(n)}{n(n+1)}
 +  \frac3n +\frac{4}{n+1} + \frac{2}{n^2} -9\right], \nonumber\\
& & {}\\
C_{3,n}^{(0),q}
&=& 1, \\
C_{3,n}^{(1),q}
&=& C_F \left[ 2S_1^2(n) - 2S_2(n) +3S_1(n)-\frac{2S_1(n)}{n(n+1)}
 +  \frac1n +\frac{2}{n+1} + \frac{2}{n^2} -9\right], \nonumber\\
& & {} \\
\nopagebreak
C_{2,n}^{(0),G}
&=& 0, \\
C_{2,n}^{(1),G}
&=& -4T_F \left[ S_1(n)\frac{n^2+n+2}{n(n+1)(n+2)}
 +  \frac1n -\frac{1}{n^2}-\frac{6}{n+1}+\frac{6}{n+2}
              \right], \nonumber\\
& & {}
\end{eqnarray}
\end{mathletters}%
with $C_F = 4/3$ and $T_F = N_f/2$.
The functions $S_1$ and $S_2$ are given in Appendix \ref{APPunp}.

The analogous expression for the moments of the polarized
structure function $g_1(x,Q^2)$ (for $N_f=3$) is:
\begin{eqnarray}
\label{g1Q}
{\cal G}_{1,n}(Q^2)
&=& \Delta C_n^{NS}(Q^2)
    \left( \pm \frac{1}{12}\Delta q_{3,n}(Q^2)
             + \frac{1}{36}\Delta q_{8,n}(Q^2)
    \right)             \nonumber\\
&+& \frac{1}{9} \Delta C_n^q(Q^2) \Delta \Sigma_{n}(Q^2)
 +  \frac{1}{9} \Delta C_n^G(Q^2) \Delta G_{n}(Q^2) ,
\end{eqnarray}
where $\pm$ refers to the proton and neutron, respectively.
The polarized Wilson coefficients $\Delta C_n$ are given by
\cite{MN95,Ko80}:
\begin{eqnarray}
\Delta C_n^{NS}(Q^2)
&=& \Delta C_n^q(Q^2)
 =  \Delta C_n^{(0), q}
 + {\alpha_s(Q^2) \over 4\pi} \Delta C_n^{(1),q},  \\
\Delta C_n^G(Q^2)
&=& \Delta C_n^{(0),G}
 + {\alpha_s(Q^2) \over 4\pi} \Delta C_n^{(1),G},
\end{eqnarray}
where
\begin{mathletters}
\begin{eqnarray}
\Delta C_n^{(0), q}
&=& 1,                                  \\
\Delta C_n^{(1), q}
&=& C_F \left[ 2S_1^2(n) - 2 S_2(n) + \frac{2 S_1(n)}{n+1}
 - \frac{2 S_1(n)}{n} + 3 S_1(n) - \frac{2}{n(n+1)}
 + \frac3n + \frac{2}{n^2} - 9 \right] ,\nonumber\\
& & {}                                  \\
\Delta C_n^{(0), G}
&=& 0,                                  \\
\Delta C_n^{(1), G}
&=& T_F \left[ \frac{4 (n-1)(1 - n - n S_1(n))}{n^2(n+1)} \right].
\end{eqnarray}
\end{mathletters}%
As in the unpolarized case, one has to use anomalous dimensions
and Wilson coefficients calculated consistently within the same
factorization scheme.
For the NLO evolution of both unpolarized and polarized distributions
we work within the $\overline{\mbox{MS}}$-scheme throughout.
A property of this scheme is that $\Delta C_{n=1}^G = 0$,
resulting in zero gluonic contribution to the first moment of $g_1$,
and hence no correction to the Ellis-Jaffe sum rule \cite{EJ}.
All other moments of the structure function $g_1(x,Q^2)$ do,
however, contain gluonic corrections.

\section{Numerical Solution}
\label{numsol}

Having outlined the essential features of the DGLAP evolution
equations, we next examine methods which can be used to efficiently
and reliably extract their solutions.
The most direct approach involves simply integrating the
integro-differential version of the equations directly
(see e.g. Ref.\cite{MK95}).
With this method, to obtain evolved distributions in the range
$[x_0,1]$ the input distribution needs to be known only for
$x>x_0$.
However, many iterations are necessary to obtain accurate results when 
evolving over a large range of $Q^2$, which consequently decreases 
both the efficiency and accuracy of the evolution. 
Other proposed methods have made use of complete sets of orthogonal
polynomials, such as Bernstein polynomials \cite{SCHRPOL} or
Laguerre polynomials \cite{KKK94}.
Difficulties exist, however, in applying these methods to
regions of very small or very large $x$.
Instead, we will follow an approach using the Mellin transform
technique \cite{Fo77}, similar to that discussed in
Ref.\cite{GR90}, but differing in a number of aspects which 
we now discuss.

\subsection{Contour Method}

The advantage of the Mellin transform (or contour integration)
technique lies in its simplicity, and efficiency with computing
time.
In the past this method has been applied in cases where the moments
of the input distributions were known in analytic form, such as when
$q(x)$ is a linear combination of terms of the form $a x^b (1-x)^c$
or similar \cite{GR90}.
The need often exists, however, to evolve input distributions which
are not able to be expressed in this form.
For example, one may wish to evolve distributions obtained from
model calculations \cite{CQ}, which generally do not have a simple  
analytic form.
Therefore it is necessary that one has available an accurate NLO
evolution procedure for applications where the input distributions 
are known only at a limited number of points in $x$.

The main steps involved in the Mellin transform technique can be
summarized as follows:

\begin{itemize}
\item The Mellin moments of the input distributions
${\cal Q}_n(\mu^2)$ are calculated numerically at the input
scale $\mu^2$.

\item The calculated moments are evolved to the scale $Q^2$ via
Eqs.(\ref{evoNS})and (\ref{evoS}).

\item The parton distributions are reconstructed via the inverse
Mellin transformation:
\begin{mathletters}
\begin{equation}
q(x,Q^2)
= \frac{1}{2\pi i} \int_{c-i\infty}^{c+i\infty} dn\
  x^{-n} {\cal Q}_n(Q^2).
\label{IMT}
\end{equation}
In practice it is more convenient to express Eq.(\ref{IMT})
as an inverse Laplace transformation:
\begin{equation}
q(x,Q^2) = \frac{1}{2\pi i} \int_{c-i\infty}^{c+i\infty}
            dn\ e^{n(-\ln x)}\ {\cal Q}_n(Q^2).
\label{ILT}
\end{equation}
\end{mathletters}%
\end{itemize}

One disadvantage in principle of this method is that in order 
to calculate the moments of distributions, the input must be 
known in the entire $x$-region $[0,1]$.
However, as will be seen in Section \ref{res}, the restriction
to the interval [$x_0,1$] has negligible effects numerically.

\subsection{Evaluation of Moments}

If the input distribution is of the form $q(x,\mu^2) = a x^b (1-x)^c$,
its moments are simply expressed in terms of Beta functions $B$:
\begin{equation}
{\cal Q}_n = a B(n+b,c+1) = a \frac{\Gamma(n+b)\Gamma(c+1)}
                            {\Gamma(n+b+c+1)}.
\end{equation}
If the analytic moments are not available, each Mellin moment must 
be integrated numerically.
This causes difficulties, however, for complex values of $n$,
since the integrand in Eq.(\ref{ILT}) oscillates rapidly as
$x \rightarrow 0$.
This can be seen by writing $n = c + i d$, where $c,d$ are real:
\begin{mathletters}
\begin{equation}
{\cal Q}_n(\mu^2) = \int_0^1 dx\ x^{-1+c} e^{id\ln x}\ q(x,\mu^2).
\end{equation}
In practice it is easier to perform the integration by making the
substitution $\ln x \rightarrow t$, which gives an integrand that
oscillates with constant frequency:
\begin{equation}
{\cal Q}_n(\mu^2) = \int_{-\infty}^0 dt\ e^{(c-1) t} e^{idt}
             q(e^t,\mu^2).
\end{equation}
\end{mathletters}%
The real and imaginary parts of this integral are then evaluated
separately by using a suitable adaptive quadrature algorithm
(such as that provided by the IMSL numerical library routine
DQDAWO \cite{IMSL}).
This way accurate moments can be obtained for any input
distribution given as a function of Bjorken-$x$ over the entire
region, $0 \leq x \leq 1$.

Two additional problems arise when using input
that is given only at a limited number of $x$-points.
Firstly, without an adaptive quadrature routine the moments are
often inaccurate, with the error increasing with the imaginary
part of $n$.
One is then forced to interpolate the input between the $x$-points,
which leads to a second problem, namely that the $x$-range covered
by these input points is limited to $x_0 \leq x \leq x_1$.

Clearly the quality of the moments, and hence the accuracy of the
evolved parton distributions, depends strongly on the quality of
the interpolation routine.
For this we have used cubic splines and higher order B-splines, 
both with a `not-a-knot' condition (see e.g. IMSL routines DCSINT 
and DBSINT, respectively \cite{IMSL}).
Taking equidistant $x$-points leads to very good results everywhere
except in the neighborhood of the boundaries at $x = x_0$ and $x_1$.
To obtain sufficiently good results over the entire $x$-range one
should choose the location of $x$-points suitably, in particular
more points near the boundaries than in between.

\subsection{Numerical Inversion}

The final step, which is numerically the most difficult, is the contour
integration in the inverse Mellin (or Laplace) transformation.
Depending on the value of Bjorken-$x$ at which the parton distribution
is to be reconstructed, two different methods are employed for the
regions $x<0.80$ and $x\ge 0.80$.

In the $x<0.80$ region Crump's method \cite{Cr76} is applied
(using e.g. the IMSL routine DINLAP \cite{IMSL}),
in which the inverse Laplace transform of the function $F(n)$
is approximated by a partial sum of a Fourier representation:
\begin{equation}
f(t)
\simeq  \frac{e^{ct}}{\tau} \left[ \frac12 F(c) + \sum_{k=1}^{\infty}
  \left\{ \Re e \left( F\left( c+k\pi i/\tau \right) \right)
  \cos \frac{k\pi t}{\tau}
  -\Im m \left( F\left( c+k\pi i/\tau \right) \right)
   \sin \frac{k\pi t}{\tau} \right\} \right].
\end{equation}
The value of $c$ has to be on the right side  of all singularities
appearing in $F(n)$.
{}From Regge physics one obtains the conditions $\Re e\ n \ge 2$
and $\Re e\ n \ge 1$ for the singlet and non-singlet distributions,
respectively.
The parameter $\tau$ controls the step length in the imaginary
direction.
This method fails, however, when $f(t)$ reaches values close to
zero, which occurs for large $x$.
In the kinematic range $x \ge 0.8$ we therefore utilize an alternative
algorithm, in which the inverse Laplace transform is computed via a 
Laguerre expansion (see e.g. IMSL routine DSINLP \cite{IMSL}).
The performance of this procedure has been checked by first performing
the Mellin transform by computing the moments, and then transforming
back with the above routines to compare the output with the original
input.

One further comment should be made about the small-$x$ region.
It is in general more problematic to compute the inverse Laplace
transform $f(t)$ numerically when $t$ is very large, or equivalently
when $x$ is very small.
In practice we find that the above procedure gives reliable results 
for parton distributions at least down to $x$ values reached by current
experiments at HERA, namely $x \sim 10^{-4}$.

\subsection{Computation of the Running Coupling Constant}

If one performs evolution at low values of $Q^2$, care must be
taken when calculating the running coupling constant $\alpha_s$.
At NLO, $\alpha_s$ can be computed exactly by solving the 
transcendental equation numerically:
\begin{equation}
\label{coupling}
 \ln\frac{Q^2}{\Lambda_{QCD}^2}
- \frac{4\pi}{\beta_0\alpha_s}
+\frac{\beta_1}{\beta_0^2}\ln\left[\frac{4\pi}{\beta_0\alpha_s}
+\frac{\beta_1}{\beta_0^2}
 \right] = 0,
\end{equation}
as obtained from the renormalization group analysis.
For small values of $\alpha_s$ one can find approximate solutions
by using the formula:
\begin{equation}
\label{alpha_appr}
 \alpha_s(Q^2)
\approx
  \frac{4\pi}{\beta_0\ln Q^2/ \Lambda_{QCD}^2}-
  \frac{4\pi\beta_1}{\beta_0^3}\frac{\ln\ln Q^2/ \Lambda_{QCD}^2}
  {\ln^2Q^2/ \Lambda_{QCD}^2}.
\end{equation}
In practice this approximation turns out to be accurate only for
$Q^2 \agt 1$ GeV$^2$.
At smaller scales, $Q^2 \approx 0.4$ GeV$^2$, the error in $\alpha_s$
is about 7\% (with $\Lambda_{QCD} = 200$ MeV in both
Eq.(\ref{coupling}) and (\ref{alpha_appr}) and $N_f = 3$).
In the region $Q^2 \approx 0.2$ GeV$^2$ from which 
low energy model calculated input is often evolved, 
use of the approximation Eq.(\ref{alpha_appr})
can give errors of up to 20\% in the running coupling
for the same values of $\Lambda_{QCD}$ and $N_f$.
The discrepancy becomes bigger still for larger $\Lambda_{QCD}$.
Of course $\Lambda_{QCD}$ itself can be extracted using
Eq.(\ref{alpha_appr}) from data at large $Q^2$, since 
there the difference between Eqs.(\ref{coupling}) and
(\ref{alpha_appr}) is minimal, however the use of
Eq.(\ref{alpha_appr}) outside this region could be problematic 
given the above differences.

\section{Results}
\label{res}

Using the numerical method described in Section III, we
present here the results of the evolution of parton densities
including all the NLO corrections.
We should stress that our aim is not to perform a complete fit to 
the available data --- such programs of data parametrizations are 
adequately addressed in Refs.\cite{MRS,CTEQ,GR90}, for example.   
Our purpose will be to simply demonstrate the potential differences 
in the evolved distributions when order $\alpha_s^2$ corrections 
are taken into account.  
After first establishing the degree of accuracy and the region of
reliability of the evolution program in the unpolarized sector, we
then turn our attention to the more topical discussion of evolution
of polarized distributions at NLO.

\subsection{Unpolarized Distributions}
\label{resUNP}

To demonstrate the effect of the numerical algorithm on a 
typical parton distribution, 
for the spin-averaged distributions we take as input the
latest parametrization from the CTEQ collaboration \cite{CTEQ}:
\begin{mathletters}
\begin{eqnarray}
xu_V(x) &=& 1.37 x^{0.497}(1-x)^{3.74}(1+6.25x^{0.880}),\label{upar}\\
xd_V(x) &=& 0.801 x^{0.497}(1-x)^{4.19}(1+1.69x^{0.375}),\label{dpar}\\
xG(x)   &=& 0.738 x^{-0.286}(1-x)^{5.31}(1+7.30x),\\
x(\bar{d}(x) + \bar{u}(x))
        &=& 0.1094 x^{-0.286}(1-x)^{8.34}(1+17.5x),
\end{eqnarray}
\end{mathletters}%
given at $Q^2 = 2.56$ GeV$^2$.
Evolution at NLO is carried out in the
$\overline{{\rm MS}}$ scheme, using $\Lambda_{QCD} = 239$ MeV
for $N_f=4$ flavors, and taking into account the charm threshold
at $Q^2 = $(1.6 GeV)$^2$.
(Although the effect of neglecting the charm threshold increases with
decreasing $Q^2$, even when evolving down to $Q^2 \sim 0.5$ GeV$^2$
it is still very small.)

To illustrate the performance of the evolution algorithm, we first
examine NLO evolution of the total valence distribution, 
$x(u_V + d_V)$.
In Fig.1 the program's accuracy is illustrated when using different
numbers of points, $N_x$, in Bjorken-$x$, with the input evolved to
$Q^2 = 100$ GeV$^2$.
With $N_x=50$ points in the range [0.005,1], the results are
indistinguishable from those obtained by calculating the
moments of (\ref{upar},\ref{dpar}) analytically.
In this case the effect of changing the lower limit of integration
from $x_0=0$ to $x_0>0$ is negligible.
Excellent results can be attained for nearly all $x$ even with 
$N_x=25$ (solid curve, marked ``$i25$'' in Fig.1).
Decreasing the number of points to $N_x=13$ has very little effect 
in the intermediate-$x$ region, $0.2 \alt x \alt 0.8$, but at larger 
$x$ the accuracy becomes somewhat worse.
For comparison, we also show in Fig.1 (dashed curves) the results
obtained by calculating moments via a simple histogram method,
\begin{eqnarray}
{\cal Q}_n(Q^2)
&=& \int_0^1 dx\ x^{n-2} \left( x q(x,Q^2) \right)   \nonumber\\
&\approx& \sum_{i=1}^{N_x} x_i q(x_i,Q^2)
          \left( { x_i^{n-1} - x_{i-1}^{n-1} \over n - 1 }
          \right).
\end{eqnarray}
For $N_x=50$ the discrepancy with the interpolation method is rather
more dramatic, particularly at large $x$, and to obtain a similar
degree of accuracy one needs $N_x$ in excess of 1000.
Similar accuracy will also be obtained when evolving different
input distributions, such as those from Refs.\cite{CQ,MRS,GR90}.

In Fig.2 we compare the valence $x (u_V + d_V)$ distribution
evolved with leading and next-to-leading order corrections.
For the LO evolution we use $\Lambda_{QCD} = 177$ MeV,  
according to the latest fit results from Ref.\cite{CTEQ}.
To avoid ambiguities associated with the factorization scheme
dependence of the (generally unphysical) NLO distributions, we
work in the DIS scheme.
(Differences between distributions calculated in different schemes
are of order $\alpha_s(Q^2)$ \cite{MT91}, and while negligible at
high $Q^2$, they can become quite large at $Q^2 \alt 1$ GeV$^2$.)
Although in general input distributions at LO and NLO need not be 
identical, especially when one seeks precisions fits to DIS data, 
we will simply demonstrate the effects by assuming the input  
shapes to be the same.  
This assumption is an implicit one, for example, in low-energy
effective model calculations of leading twist parton distributions
\cite{JR,CQ}, 
where because the input model scale is {\em a priori} unknown, 
the same calculated distribution must be evolved with either 
LO or NLO corrections.

The main effect of NLO corrections on the valence distributions
(with upwards evolution) is to make them softer compared to those
evolved with LO corrections only.
Conversely, NLO evolution produces somewhat harder distributions
when evolved downwards to low scales $Q^2 = 0.5$ GeV$^2$, such as
those associated with valence distributions calculated from
low-energy (constituent quark) models of QCD.

In connection with this, we should note that uncertainty in the
precise value of $\Lambda_{QCD}$ (the Particle Data Group gives
$234 \pm 56$ MeV \cite{PDG}) can be translated into quite a significant
difference in the starting scale $Q_0^2$ for the evolution of the 
calculated model results.
Taking the upper and lower values of $\Lambda_{QCD} = 290$ and 178 MeV,
one obtains essentially identical shapes of the evolved distributions
with $Q_0^2 = 1$ and 0.5 GeV$^2$, respectively.
Even given the fact that the exact scale for such model calculations
is a priori unknown, a factor 2 (!) represents quite a large effect
in the phenomenology of the scale dependence of low-energy model
distributions.

Turning now to the singlet sector, we illustrate in Fig.3 the effects
of NLO corrections on the distribution $\Sigma(x,Q^2)$.
Taking the CTEQ parametrization at $Q_0^2 = 2.56$ GeV$^2$ as input,
the distribution is evolved to $Q^2 = 10$ and 100 GeV$^2$ using the
DIS scheme to compare physical quantities at LO and NLO (in the DIS
scheme $x\Sigma(x,Q^2)$ represents the structure function $F_2$
measured in deep-inelastic neutrino-nucleon scattering).
The results here have been obtained with $N_x=60$ on a logarithmic
scale in Bjorken-$x$, and agree to within 0.01\% with those using
analytic moments.
Similar accuracy is obtained also for the singlet gluon distribution.

Having established the quality of performance of our numerical method
for both non-singlet and singlet evolution of unpolarized distributions,
in the next Section we discuss the NLO effects on polarized parton
densities.

\subsection{Polarized Distributions}
\label{polRES}

Evolution of polarized parton distributions in the non-singlet sector
is straightforward, since at NLO it is governed by exactly the same
anomalous dimensions as for unpolarized DIS.
Therefore the effects are identical to those in Figs.1 and 2.
In the singlet sector, on the other hand, the anomalous dimensions
for polarized and unpolarized scattering are very different.
In fact, due to the complexity of the problem, the relevant splitting 
functions and anomalous dimensions necessary to describe the NLO 
singlet evolution as a function of $x$ have only recently been 
calculated \cite{MN95}.
A recalculation by Vogelsang \cite{Vo95}, using an independent method,
has confirmed the results in \cite{MN95}.
To use the inverse Mellin method for the evolution, it is necessary 
to analytically continue the results of Ref.\cite{MN95} into the 
complex-$n$ plane.
In doing so care must be taken since, unlike for the unpolarized
distributions, the continuations for spin-dependent structure
functions are performed from odd moments.
The full details of the continuation can be found in Appendix 
\ref{APPpol}
\footnote{While completing this paper, similar NLO analyses of
polarized distributions in Refs.\cite{GRSV,BALL} came to our attention.
A different prescription for continuation of the singlet  
polarized anomalous dimensions into complex-$n$ space is used  
in Ref.\cite{GRSV}, although we have checked that numerically    
these are in fact equivalent --- see Appendix \ref{APPpol}.}.

To illustrate the role of NLO corrections in spin-dependent
distributions, we have used as input the global fit by Gehrmann
and Stirling (Set A) \cite{GS95}:
\begin{mathletters}
\begin{eqnarray}
x\Delta u_V(x)
&=& 0.327 x^{0.46} (1-x)^{3.64} (1+ 18.36 x), \\
x\Delta d_V(x)
&=& -0.139 x^{0.46} (1-x)^{4.64} (1+ 18.36 x), \\
x\Delta \overline{u}(x)
&=& x\Delta \overline{d}(x)\
 =\ x\Delta \overline{s}(x)\
 =\ x\Delta c(x)\
 =\ 0, \\ 
x\Delta G(x) &=& 16.64 \,x \,(1-x)^{7.44},
\end{eqnarray}
\end{mathletters}%
with a starting scale of $Q^2 = 4$ GeV$^2$.
Following Ref.\cite{GS95}, in LO we again perform the evolution with 
three active flavors and $\Lambda^{LO}_{QCD} = 177$ MeV, as for the 
unpolarized evolution above.
However, contrary to the procedure adopted in \cite{GS95}, we omit
the anomalous gluon term when calculating the structure functions
$g_1^p(x,Q^2)$ and $g_1^n(x,Q^2)$ at leading order.
This term arises only at NLO, and in a consistent treatment should
only be included in NLO evolution --- see Eqs.(\ref{g1con}) and
(\ref{g1Q}).
In NLO we also use three active flavors, 
and a different value of
$\Lambda^{NLO}_{QCD} = 239$ MeV, corresponding to that obtained
by CTEQ in their fits to unpolarized parton distributions.
As in the discussion of the unpolarized evolution in Sec.IV A, 
for illustration purposes we take the same input shape for both 
LO and NLO evolution.

In Fig.4 the $x$ dependence of $\Delta\Sigma(x,Q^2)$, evolved down 
to 2 GeV$^2$ and up to 10 GeV$^2$, is shown between $x=10^{-4}$ and 1.
The difference between distributions evolved in LO and NLO   
can be sizeable below 
$x \approx 5 \times 10^{-3}$, and even larger 
with decreasing $x$.
Furthermore, it is interesting to observe that at $Q^2 = 10$ GeV$^2$ 
even the
curvature changes sign compared to the LO result.
Similarly large differences between LO and NLO evolution appear for
the polarized gluon distribution, $\Delta G(x,Q^2)$, at small $x$ 
--- see Fig.5.
These dramatic effects are largely independent of the shapes of the
input LO and NLO distributions, and  
can be traced back to the dominant role played 
by NLO corrections at small $x$.
This becomes evident if one considers the $x \rightarrow 0$ behavior 
of the splitting functions
$P_{ij}(x) = P_{ij}^{(0)}(x)
           + \left( \alpha_s /4 \pi \right) P_{ij}^{(1)}(x)$ 
\cite{MN95,HN91} --- in LO one has:
\begin{mathletters}
\begin{eqnarray}
P_{qq}^{(0)}(x \rightarrow 0)  
& \rightarrow & 4 C_F, \\
P_{qG}^{(0)}(x \rightarrow 0)  
& \rightarrow & - 8 T_F , \\
P_{Gq}^{(0)}(x \rightarrow 0)  
& \rightarrow & 8 C_F ,\\
P_{GG}^{(0)}(x \rightarrow 0)  
& \rightarrow & 16 C_A ,
\end{eqnarray}
while at NLO:
\begin{eqnarray}
P_{qq}^{(1)}(x \rightarrow 0)  
& \rightarrow & \left( 8 C_F C_A -16 C_F T_F  - 12 C_F^2 
                \right) \ln^2 x,\\
P_{qG}^{(1)}(x \rightarrow 0)  
& \rightarrow & - \left( 16 C_A T_F  + 8 C_F T_F 
                  \right) \ln^2 x,\\
P_{Gq}^{(1)}(x \rightarrow 0)  
& \rightarrow & \left( 16 C_A C_F + 8 C_F^2 \right) \ln^2 x,\\
P_{GG}^{(1)}(x \rightarrow 0)  
& \rightarrow & \left( 32 C_A^2 -16 C_F T_F 
                \right) \ln^2 x.
\end{eqnarray}
\end{mathletters}%
For small enough $x$, the NLO corrections $P_{ij}^{(1)}(x)$ can 
become quite large and overwhelm the $\alpha_s / 4\pi$ factor. 
Note also that even the {\em sign} of $P_{qq}^{(1)}(x)$ changes
compared with $P_{qq}^{(0)}(x)$ as $x \rightarrow 0$
(for $C_F=4/3, T_F=N_f/2, C_A=3$ and $N_f \ge 2$).

With the $Q^2$ evolution of the polarized quark singlet and gluon
distributions known, one can now examine the effect of NLO corrections
on the spin-dependent structure function $g_1(x,Q^2)$.
In Fig.6 we show the proton structure function $x g_1^p(x,Q^2)$,
evolved to the averaged scale $Q^2 = 10$ GeV$^2$ at which the
SMC data \cite{SMCP} were taken.
The important result to notice is that the NLO corrections are
large in the range of $x$ covered by the experiment.
In the region $x \sim 0.2$ the corrections can in fact be of the
same order of magnitude as the quoted error bars on the data.
Furthermore, the fact that these are negative is consistent with 
the known reduction of the first moment of $g_1$ \cite{Q2MOM}
when NLO corrections are included.

In Fig.7 we also show the polarized neutron structure function
$x g_1^n(x,Q^2)$, evolved at LO and NLO down to a scale
$Q^2 = 2$ GeV$^2$ for comparison with the SLAC E142 data \cite{E142}.
Because evolution here involves a somewhat lower resolution scale, 
the effects of NLO corrections should be be relatively larger.
Clearly, one sees that these can have a potentially significant 
effect on the shape of the structure function, when the data points 
are evolved to the same value of $Q^2$.

\section{Conclusion}

We have examined in detail the role of NLO corrections on the shapes 
of the nucleon's singlet and non-singlet parton distributions, using 
a $Q^2$ evolution algorithm based on the inverse Mellin transform in
complex moment space.  
For a complete description of polarized and unpolarized evolution at 
NLO it has been necessary to extrapolate recently obtained results 
for the polarized singlet NLO anomalous dimensions into the complex-$n$
plane.
Of particular importance are the $Q^2$ corrections to the 
spin-dependent $g_1$ structure functions of the proton and neutron.
We find that in the region of $x$ and $Q^2$ relevant to current and
upcoming experiments, the NLO effects are not negligible, and should
be included in future analyses of data.
Finally, we should make a note about the $Q^2$ behavior of the 
polarization asymmetry $A_1 = g_1/F_1$, which was assumed to be
independent of $Q^2$ in previous data analyses \cite{SMCP,E142}.
While the non-singlet parts of $g_1$ and $F_1$ evolve 
similarly at NLO, the evolution of the singlet components is 
dramatically different, as seen in Figs.3--5, especially at
low $x$.
Therefore one should be wary of the possible errors introduced
into $g_1$ at small $x$ if one extracts this from the $A_1$ data
under this assumption.

\acknowledgements

We thank S.Kumano, E.Reya, F.M.Steffens and A.Vogt 
for useful discussions 
and communications, and in particular W.Weise for helpful comments 
and a careful reading of the manuscript.
We also thank A.W.Schreiber for providing the LO evolution program.
This work was supported by the BMBF.

\newpage

\appendix
\section{Anomalous Dimensions}
\label{APP}

\subsection{Unpolarized DIS}
\label{APPunp}

Although the results for the full anomalous dimensions
up to ${\cal O}(\alpha_s^2)$ can be found throughout the 
literature, for convenience we catalogue them here in full.
In the $\overline{\mbox{MS}}$ scheme the anomalous dimensions are
given by the perturbative expansion:
\begin{mathletters}
\begin{eqnarray}
\gamma_{NS}^n(Q^2)
& = &  \frac{\alpha_s(Q^2)}{4\pi}\gamma_{NS}^{(0),n}
  +  \left(\frac{\alpha_s(Q^2)}{4\pi}\right)^2 \gamma_{NS}^{(1),n}
+\cdots,      \\
\mbox{\boldmath$\gamma$}^n(Q^2)
& = & \frac{\alpha_s(Q^2)}{4\pi} \mbox{\boldmath$\gamma$}^{(0),n}
  + \left(\frac{\alpha_s(Q^2)}{4\pi}\right)^2
    \mbox{\boldmath$\gamma$}^{(1),n}
+  \cdots,     \\
\mbox{\boldmath$\gamma$}^{(k),n} & = & \left(
        \begin{array}{cc} \gamma_{qq}^{(k),n} & \gamma_{qG}^{(k),n}\\
              {}              &         {}            \\
        \gamma_{Gq}^{(k),n} & \gamma_{GG}^{(k),n} \\
        \end{array}  \right),
\end{eqnarray}
\end{mathletters}%
where the order $\alpha_s$ anomalous dimensions are \cite{AP77,FK81}:
\begin{mathletters}
\label{gam0unp}
\begin{eqnarray}
\gamma_{NS}^{(0),n}
& = & \gamma_{qq}^{(0),n}  =  2 \, C_F \, \left[ 4 S_1(n) -3 -
   \frac{2}{n(n+1)}\right], \\
\gamma_{qG}^{(0),n}
& = & -8 \, T_F \, \frac{n^2+n+2}{n(n+1)(n+2)},\\
\gamma_{Gq}^{(0),n}
& = & -4 \, C_F \, \frac{n^2+n+2}{(n-1)n(n+1)},\\
\gamma_{GG}^{(0),n}
& = & 2 \, C_A \, \left[4S_1(n)-\frac{11}{3} -\frac{4}{n(n-1)}
    - \frac{4}{(n+1)(n+2)}\right] +\frac83 T_F ,
\end{eqnarray}
\end{mathletters}%
with $C_F=4/3$, $T_F=N_f/2$ and $C_A = 3$.
The anomalous dimensions at ${\cal O}(\alpha_s^2)$ are
given by \cite{FK81}:
\begin{mathletters}
\label{gam1unp}
\begin{eqnarray}
\gamma_{NS}^{(1),n} & = & C^2_F \left[ \frac{16 S_1(n) (2n+1)}
{n^2  (n+1)^2} +16 \left( 2 S_1(n) - \frac{1}{n(n+1)} \right)
\left( S_2(n) - S_2'(n/2) \right) \right. \nonumber\\
& &  + 24 S_2(n) + 64 \tilde{S}(n)
 - 8 S_3'(n/2) -3\nonumber\\
& & \left. - 8 \frac{3n^3+n^2-1}
{n^3(n+1)^3} - 16 \eta\frac{2n^2+2n+1}
{n^3(n+1)^3} \right] \nonumber\\
& & + C_A C_F \left[ \frac{536}{9} S_1(n)
 - 8 \left( 2 S_1(n)- \frac{1}{n(n+1)} \right) (2 S_2(n)
 - S_2'(n/2)) \right. \nonumber\\
& & - \frac{88}{3} S_2(n) - 32 \tilde{S}(n)
 + 4 S_3'(n/2) - \frac{17}{3} \nonumber\\
& & \left. - \frac49 \frac{151n^4+236n^3+88n^2+3n+18}
{n^3 (n+1)^3} + 8 \eta \frac{2n^2+2n+1}
{n^3 (n+1)^3} \right] \nonumber\\
& & + C_F T_F 
\left[ -\frac{160}{9} S_1(n)+\frac{32}{3} S_2(n)+\frac43 +
\frac{16}{9} \frac{11n^2+5n-3}{n^2 (n+1)^2} \right], \nonumber\\
& & {} \\
\gamma_{qq}^{(1),n} & = & \gamma_{NS}^{(1),n}(\eta = 1)
- 16 C_F T_F  \frac{5n^5+32n^4+49n^3+38n^2+28n+8}
{(n-1)n^3(n+1)^3(n+2)^2}, \nonumber\\
& & {}
\end{eqnarray}

\begin{eqnarray}
          \gamma_{qG}^{(1),n} & = & -8 C_A T_F  \left[ (-2 S_1^2(n)
          +2 S_2(n)-2 S_2'(n/2) ) \frac{n^2+n+2}
          {n(n+1)(n+2)} \right. \nonumber\\
          & & + \frac{8 S_1(n)(2n+3)}
          {(n+1)^2(n+2)^2}
          + 2 \frac{n^9+6n^8
          +15n^7+25n^6+36n^5+85n^4}{(n-1)n^3(n+1)^3(n+2)^3} \nonumber\\
          & & \left. +2 \frac{128n^3+104n^2+64n+16}
          {(n-1)^3n^3(n+1)^3(n+2)^3} \right] \nonumber\\
          & & -8 C_F T_F  \left[
          (2 S_1^2(n)-2 S_2(n)+5) \frac{n^2+n+2}
          {n(n+1)(n+2)} - \frac{4 S_1(n)}{n^2} \right. \nonumber\\
          & & \left. + \frac{11n^4+26n^3+15n^2+8n+4}
          {n^3(n+1)^3(n+2)} \right], \nonumber\\
          & & {}
\end{eqnarray}

\begin{eqnarray}
          \gamma_{Gq}^{(1),n}
          & = & -4C^2_F \left[ (-2 S_1^2(n)+10 S_1(n)
          -2 S_2(n)) \frac{ n^2+n+2}{n-1)n(n+1)}
          -4 \frac{S_1(n)}{(n+1)^2} \right. \nonumber\\
          & & \left. -\frac{12n^6+30n^5+43n^4
          +28n^3-n^2-12n-4}{(n-1)n^3(n+1)^3} \right] \nonumber\\
          & & -8 C_A C_F \left[ (S_1^2(n)+S_2(n)-S_2'(n/2))
          \frac{n^2+n+2}{(n-1)n(n+1)} \right. \nonumber\\
          & & - S_1(n) \frac{17n^4+41n^2-22n-12}{3(n-1)^2
          n^2(n+1)}+ \frac{n^3+n^2+4n+2}
          {n^3(n+1)^3} \nonumber\\
          & & + \frac{109n^8+512n^7
          +879n^6+772n^5-104n^4}{9(n-1)^2n^3(n+1)^2(n+2)^2}\nonumber\\
          & & \left.
              \frac{-954n^3-278n^2+288n+72}{9(n-1)^2n^3(n+1)^2(n+2)^2}
          \right] \nonumber\\
          & & - \frac{32}{3} C_F T_F \left[ \left(S_1(n)- \frac83\right)
          \frac{n^2+n+2}{(n-1)n(n+1)} + \frac{1}{(n+1)^2} \right],
          \nonumber\\
          & & {}
\end{eqnarray}

\begin{eqnarray}
          \gamma_{GG}^{(1),n}
          & = & C_A T_F  \left[ -\frac{160}{9} S_1(n)
          +\frac{32}{3} +\frac{16}{9} \frac{38n^4+76n^3+94n^2+56n+12}
          {(n-1)n^2(n+1)^2(n+2)} \right] \nonumber\\
          & & + C_F T_F \left[ 8+16 \frac{2n^6+4n^5
          +n^4-10n^3-5n^2-4n-4}{(n-1)n^3
          (n+1)^3(n+2)} \right] \nonumber\\
          & &+ C^2_A \left[ \frac{536}{9} S_1(n)+64S_1(n) \frac{2n^5
          +5n^4+8n^3+7n^2-2n-2}
          {(n-1)^2n^2(n+1)^2(n+2)^2} - \frac{64}{3} \right. \nonumber\\
          & & +32 S_2'(n/2) \frac{n^2+n+1}{(n-1)n(n+1)
          (n+2)} \nonumber\\
          & & -\frac49 \frac{457n^9
          +2742n^8+6040n^7+6098n^6+1567n^5
          -2344n^4-1632n^3}{(n-1)^2n^3(n+1)^3(n+2)^3}\nonumber\\
          & & \left. + \frac{560n^2+1488n+576}
          {(n-1)^2n^3(n+1)^3(n+2)^3}- 16 S_1(n) S_2'(n/2)
           + 32 \tilde{S}(n)
          - 4 S_3'(n/2) \right]. \nonumber\\
          & & {}
\end{eqnarray}
\end{mathletters}%

The expressions (\ref{gam0unp}) and (\ref{gam1unp}) are also valid 
for complex $n$
if the following analytic continuations are used \cite{GR90}:
\begin{mathletters}
\label{Sunp}
\begin{eqnarray}
S_1(n) & = & \sum_{j=1}^n \frac1j \to \gamma_E+\Psi(n+1),\\
S_2(n) & = & \sum_{j=1}^n \frac{1}{j^2} \to \zeta(2) - \Psi'(n+1),\\
S_3(n) & = & \sum_{j=1}^n \frac{1}{j^3} \to \zeta(3)+\Psi''(n+1),\\
S'_l(n) & = & 2^{l-1} \sum_{j=1}^n \frac{1+(-1)^j}{j^l} \nonumber\\
      &\to& \frac12(1+\eta)S_l(n/2)+\frac12(1-\eta)S_l((n-1)/2),
          \quad l=1,2,3, \\
\widetilde{S}(n) & = & \sum_{j=1}^n \frac{(-1)^j}{j^2} S_1(j) \nonumber\\
    &\to& -\frac58\zeta(3) +\eta\left\{ \frac{S_1(n)}{n^2}
        -\frac{\zeta(2)}{2}\left[\Psi((n+1)/2) -\Psi(n/2)\right]
        \right. \nonumber \\
    & &  \left. \hspace*{1cm}
         +\int_0^1 dx x^{n-1} \frac{\mbox{Li}_2(x)}{1+x}\right\},
\end{eqnarray}
\end{mathletters}%
where
\begin{mathletters}
\begin{eqnarray}
\gamma_E & = & 0.577216,\\
\zeta(2) & = & \pi^2/6,\\
\zeta(3) & = & 1.202057,
\end{eqnarray}
\end{mathletters}%
and the Polygamma functions, $\Psi$, are defined as
\begin{equation}
\Psi^{(m)}(n) = \frac{d^{(m+1)} \Gamma(n)}{dn^{m+1}}.
\end{equation}
The integral over the Dilogarithm $\mbox{Li}_2(x)$ in $\widetilde{S}(n)$
can be calculated analytically by using the approximation:
\begin{equation}
\label{Lifit}
 \frac{\mbox{Li}_2(x)}{1+x} \simeq
 0.0030 + 1.0990 x -1.5463 x^2 + 3.2860 x^3 - 3.7887 x^4 + 1.7646 x^5,
\end{equation}
obtained from a least squares fit.
Similar to the fit in Ref.\cite{GR90}, it is a polynomial of degree
five, but gives slightly better values
for the anomalous dimensions, as can be seen by checking the following 
relations
for the $n=2$ moments:
\begin{mathletters}
\begin{eqnarray}
 \gamma_{qq}^{(1),n=2} & = & - \gamma_{Gq}^{(1),n=2}, \\
 \gamma_{qq}^{(1),n=2} & = & - \gamma_{Gq}^{(1),n=2}.
\end{eqnarray}
\end{mathletters}%

The Polygamma functions must be computed numerically.
For the case $|n| > 10$ one applies the asymptotic expansions:
\begin{mathletters}
\begin{eqnarray}
\Psi(n) & \simeq & \ln n -\frac{1}{2n} - \frac{1}{12n^2} 
       + \frac{1}{120n^4}
       - \frac{1}{256n^6},\\
\Psi'(n) & \simeq & \frac{1}{n} + \frac{1}{2n^2} + \frac{1}{6n^3}
       - \frac{1}{30n^5} + \frac{1}{42n^7} - \frac{1}{30n^9},\\
\Psi''(n) & \simeq & - \frac{1}{n^2} - \frac{1}{n^3} - \frac{1}{2n^4}
       + \frac{1}{6n^6} - \frac{1}{6n^8} + \frac{3}{10n^{10}}
       - \frac{5}{6n^{12}}.
\label{Psi2}
\end{eqnarray}
To obtain greater accuracy for $\Psi''(n)$ we apply an Euler
transformation to the asymptotic series in (\ref{Psi2}), ending
up with:
\begin{equation}
\Psi''(n)  \simeq  - \frac{1}{n^2} - \frac{1}{n^3} - \frac{1}{2n^4}
          + \frac{1}{6n^6} -\frac{1}{16n^8} - \frac{3}{20n^{10}}
          - \frac{5}{48n^{12}}.
\end{equation}
\end{mathletters}%
In the case of $|n| < 10$ the recursion formula
\begin{equation}
\Psi^{(m)}(n+1) = \Psi^{(m)}(n) + (-1)^m m!\ n^{1-m}
\end{equation}
is utilized repeatedly \cite{GR90}.

At next-to-leading order, a factor $(-1)^n$ appears in the anomalous 
dimensions \cite{RS79} (Eq.(\ref{Sunp})).
Therefore the analytic continuation depends on whether the evolution 
equations for a particular parton distribution are valid for even or 
odd moments, in which case:
\begin{equation}
  (-1)^n \longrightarrow \eta=\pm 1,
\end{equation}
for $n$ even or odd, respectively.
The $\eta$ factors relevant for various combinations of parton 
distributions can be derived from the crossing relations:   
\begin{mathletters}
\begin{eqnarray}
F_1(-x,Q^2)     & = & +F_1(x,Q^2),      \\
F_{2,3}(-x,Q^2) & = & -F_{2,3}(x,Q^2).
\end{eqnarray}
\end{mathletters}%
For the non-singlet combinations these are (for $N_f = 4$ active 
flavors) \cite{GR90}:
\begin{mathletters}
\begin{eqnarray}
&& u-\bar{u} \quad; \eta =-1\\
&& d-\bar{d} \quad; \eta =-1\\
&& (u+\bar{u})-(d+\bar{d}) \quad; \eta =+1\\
&& (u+\bar{u})+(d+\bar{d})-2(s+\bar{s}) \quad; \eta =+1\\
&& (u+\bar{u})+(d+\bar{d})+s+\bar{s}-3(c+\bar{c}) \quad; \eta =+1.
\end{eqnarray}
\end{mathletters}%
The singlet distributions are evolved with $\eta=+1$.

\subsection{Polarized DIS}
\label{APPpol}

For completeness, we also list the full anomalous dimensions
relevant for polarized distributions at NLO.
The order $\alpha_s$ anomalous dimensions are given by \cite{AP77,MN95}:
\begin{mathletters}
\label{gam0p}
\begin{eqnarray}
\gamma_{NS}^{(0),n}
& = & \gamma_{qq}^{(0),n}  =  2 \, C_F \, \left[ 4 S_1(n) - 3 -
   \frac{2}{n(n+1)}\right], \\
\gamma_{qG}^{(0),n}
& = &  8 \, T_F \, \frac{1-n}{n(n+1)},\\
\gamma_{Gq}^{(0),n}
& = & -4 \, C_F \, \frac{n+2}{n(n+1)},\\
\gamma_{GG}^{(0),n}
& = & 2 \,  C_A \, \left[4 S_1(n) - \frac{11}{3} - \frac{8}{n(n+1)}
    \right] +\frac83 \, T_F .
\end{eqnarray}
\end{mathletters}%
Concerning the order $\alpha_s^2$,
the anomalous dimension $\gamma_{NS,qq}^{(1),n}$ for the non-singlet
operator is the same as in the unpolarized case.
The singlet dimensions, as evaluated in Ref.\cite{MN95},
are as follows
\footnote{Note the correction to the  
$\gamma_{qG}^{(1),n}, \gamma_{Gq}^{(1),n}$ anomalous 
dimensions in \cite{MN95}.}
\begin{mathletters}
\label{gam1pol}
\begin{equation}
\gamma_{qq}^{(1),n} = \gamma_{NS,qq}^{(1),n} + \gamma_{PS,qq}^{(1),n},
\end{equation}
\begin{eqnarray}
\gamma_{PS,qq}^{(1),n}
&=&
16 \,
C_F T_F  \,
\left[
 \frac{2}{(n+1)^3}
+ \frac{3}{(n+1)^2}
+ \frac{1}{(n+1)}
+ \frac{2}{n^3}
- \frac{1}{n^2}
- \frac{1}{n}
\right],
\label{aqq1}
\end{eqnarray}

\begin{eqnarray}
\gamma_{qG}^{(1),n}
&=&
16 \,
C_A T_F \;
\left[
-  \frac{S_1^2(n-1)}{n}
+  \frac{2\, S_1^2(n-1)}{n+1}
-  \frac{2\, S_1(n-1)}{n^2}
+ \frac{4 \, S_1(n-1)}{(n+1)^2}
\right.
\nonumber\\
&&
\phantom{ C_A T_F \; 16 \, }
- \frac{ S_2(n-1)}{n}
+ \frac{2\,  S_2(n-1)}{n+1}
- \frac{2 \, \widetilde{S}_2(n-1)}{n}
+ \frac{4 \, \widetilde{S}_2(n-1)}{n+1}
\nonumber\\
&&
\left.
\phantom{ C_A T_F \; 16 \, }
- \frac{4}{n}
+ \frac{3}{n+1}
- \frac{3}{n^2}
+ \frac{8}{(n+1)^2}
+ \frac{2}{n^3}
+\frac{12}{(n+1)^3}
\right]
\nonumber\\
&&
+  \,
8 \, C_F T_F \;
\left[
\frac{2 \, S_1^2(n-1)}{n}
- \frac{4 \, S_1^2(n-1)}{n+1}
- \frac{2 \, S_2 (n-1)}{n}
+\frac{4 \, S_2(n-1)}{n+1}
\right.
\nonumber\\
&&
\phantom{+ \, C_F T_F \;8 \, \;}
\left.
+ \frac{14}{n}
- \frac{19}{n+1}
- \frac{1}{n^2}
- \frac{8}{(n+1)^2}
- \frac{2}{n^3}
+ \frac{4}{(n+1)^3}
\right],
\label{aqg1}
\end{eqnarray}

\begin{eqnarray}
\gamma_{Gq}^{(1),n}
&=&
8 \, C_A C_F \;
\left[
-\frac{2\, S_1^2(n-1)}{n}
+\frac{S_1^2(n-1)}{n+1}
+\frac{16\, S_1(n-1)}{3 n}
- \frac{5 S_1(n-1)}{3( n+1)}
\right.
\nonumber\\ && \phantom{ C_A C_F \,\, }
+ \frac{2 S_2(n-1)}{n}
-\frac{S_2(n-1)}{n+1}
+\frac{4\widetilde{S}_2(n-1)}{n}
-\frac{2\widetilde{S}_2(n-1)}{n+1}
-\frac{56}{9 n}
\nonumber\\ && \phantom{ C_A C_F \,\, }
\left.
-\frac{20}{9 (n+1)}
+\frac{28}{3 n^2}
-\frac{38}{3 (n+1)^2}
-\frac{4}{n^3}
- \frac{6}{(n+1)^3}
\right]
\nonumber\\
&&
+ 4 \, C_F^2 \;
\left[
 \frac{4 \, S_1^2(n-1)}{n}
-\frac{2 \, S_1^2(n-1)}{n+1}
-\frac{8 \, S_1(n-1)}{n}
+\frac{2 \, S_1(n-1)}{n+1}
\right.
\nonumber\\ && \phantom{+ \, C_F^2 \; 4 \, }
+\frac{8 \, S_1(n-1)}{n^2}
-\frac{4 \, S_1(n-1)}{(n+1)^2}
+\frac{4 \, S_2(n-1)}{n}
-\frac{2 \, S_2(n-1)}{n+1}
\nonumber\\ && \phantom{+ \, C_F^2 \; 4 \,}
\left.
+\frac{15}{n}
-\frac{6}{n+1}
-\frac{12}{n^2}
+ \frac{3}{(n+1)^2}
+ \frac{4}{n^3}
- \frac{2}{(n+1)^3}
\right]
\nonumber\\
&&
+ 32 \, C_F T_F \;
\left[
- \frac{2 \, S_1(n-1)}{3 \, n}
+ \frac{S_1(n-1)}{3 \, (n+1)}
+ \frac{7}{9 \, n}
\right.
\nonumber\\ && \phantom{+ C_F T_F \;32  \,}
\left.
- \frac{2}{9 \, (n+1)}
- \frac{2}{3 \, n^2}
+ \frac{1}{3 \, (n+1)^2}
\right],
\label{agq1}
\end{eqnarray}

\begin{eqnarray}
\gamma_{GG}^{(1),n}
&=&
4 \, C_A^2
\left[
\frac{134}{9} \, S_1(n-1)
+ \frac{8  \, S_1(n-1)}{n^2}
- \frac{16  \, S_1(n-1)}{(n+1)^2}
\right.
\nonumber\\ && \phantom{C_A^2 \, 4 \,}
+ \frac{8  \, S_2(n-1)}{n}
- \frac{16  \, S_2(n-1)}{n+1}
+ 4 \, S_3(n-1)
\nonumber\\ && \phantom{C_A^2 \, \,}
- 8  \, S_{1,2}(n-1)
- 8  \, S_{2,1}(n-1)
+ \frac{8  \,\widetilde{S}_2(n-1)}{n}
- \frac{16 \,\widetilde{S}_2(n-1)}{n+1}
\nonumber\\ && \phantom{C_A^2 \, \,}
+ 4 \, \widetilde{S}_3(n-1)
- 8  \, \widetilde{S}_{1,2}(n-1)
- \frac{107}{9 \, n}
+ \frac{241}{9 \, (n+1)}
\nonumber\\ && \phantom{C_A^2 \, \,}
\left.
+ \frac{58}{3 \, n^2}
- \frac{86}{3 \, (n+1)^2}
- \frac{8}{n^3}
- \frac{48}{(n+1)^3}
- \frac{16}{3}
\right]
\nonumber\\
&&
+ 32 \, C_A T_F \,
\left[
\frac{-5 \, S_1(n-1)}{9} + \frac{14}{9 n} -
\frac{19}{9 \, (n+1)} - \frac{1}{3 \, n^2} -
\frac{1}{3 \, (n+1)^2} + \frac{1}{3}
\right]
\nonumber\\
&& +8 \, C_F T_F \,
\left[
-\frac{10}{n+1} + \frac{2}{(n+1)^2} + \frac{4}{(n+1)^3}
+ 1 + \frac{10}{n} - \frac{10}{n^2} + \frac{4}{n^3}
\right].
\label{agg1}
\end{eqnarray}
\end{mathletters}%

To make use of the anomalous dimensions in the contour evolution method,
they must be analytically continued into the complex-$n$ plane.
An important fact to observe is that due to the crossing relation:
\begin{eqnarray}
g_1(-x,Q^2) &=& - g_1(x,Q^2)
\end{eqnarray}
the OPE relates only {\em odd} moments of the structure function 
$g_1(x,Q^2)$ to the sum of possible twist-two operator matrix 
elements between nucleon states (each multiplied by the 
appropriate moments of the Wilson coefficient functions).
This is contrary to the case of $F_2(x,Q^2)$, where an analogous
relation holds for $n$ {\em even}.
To evolve $\Delta\Sigma(x,Q^2)$, $\Delta q_3(x,Q^2)$ and
$\Delta q_8(x,Q^2)$ one therefore has to continue the anomalous
dimensions from odd $n$.

In addition to the functions $S_{1,2,3}$, $S'_{1,2,3}$ and
$\widetilde{S}$ for the unpolarized anomalous dimensions,
for polarized DIS one must also determine the analytic continuation 
of the functions:
\begin{eqnarray}
\label{Spol}
\widetilde{S}_k(n)
&=& \sum_{j=1}^n \frac{(-1)^j}{j^k} ,\\
S_{k,l}(n)
&=& \sum_{j=1}^n \frac{S_l(j)}{j^k} ,\\
\widetilde{S}_{k,l}(n)
&=& \sum_{j=1}^n \frac{\widetilde{S}_l(j)}{j^k}.
\end{eqnarray}
Specifically, this must be done for the case of
$\widetilde{S}_1$, $\widetilde{S}_2$, $\widetilde{S}_3$,
$S_{1,2}$, $S_{2,1}$, and $\widetilde{S}_{1,2}$.
Because all these functions appear with argument $(n-1)$ in
Eq.(\ref{gam1pol}) (see below), one must continue them into
complex $n$ starting from even integers $n$.
Using the relation:
\begin{equation}
 \sum_{j=1}^n \frac{1+(-1)^j}{j^k}
= \frac{1}{2^k} \left[ (1+\eta)
  S_k\left(\frac{n}{2}\right)
   + (1-\eta)S_k\left(\frac{n-1}{2}\right)
  \right]
\label{sfun}
\end{equation}
with $\eta = + 1 (-1)$ for $n$ even (odd), a
straightforward calculation leads to:
\begin{equation}
  \widetilde{S}_k(n) = \frac{1}{2^{k-1}} S_k(n/2) - S_k(n)
\end{equation}
for $n$ even.
The equality \cite{DD84}
\begin{equation}
\frac{S_1(j)}{j^2}
= \frac{\zeta(2)}{j} - \int_0^1 dx \,x^{j-1} \mbox{Li}_2(x)
\end{equation}
then gives:
\begin{equation}
  S_{2,1}(n)
= \zeta(2) S_1(n) - \int_0^1 dx \, \frac{x^n - 1}{x-1} \mbox{Li}_2(x).
\end{equation}
The function $S_{1,2}(n)$ can be obtained from the above expressions
via \cite{GL79,DD84}:
\begin{equation}
  S_{1,2}(n) = S_1(n) S_2(n) + S_3(n) - S_{2,1}(n).
\end{equation}
The most difficult function to continue is $\widetilde{S}_{1,2}(n)$.
For this we start with (see Eq.(\ref{sfun})):
\begin{eqnarray}
\widetilde{S}_{1,2}(n) & = & \frac14 \sum_{j=1}^n \frac{S_2(j/2)}{j}
     + \frac14 \sum_{j=1}^n \frac{(-1)^j S_2(j/2)}{j}  \nonumber\\
  & + & \frac14 \sum_{j=1}^n \frac{S_2((j-1)/2)}{j}
     - \frac14 \sum_{j=1}^n \frac{(-1)^j S_2((j-1)/2)}{j}
     - \sum_{j=1}^n \frac{S_2(j)}{j},
\end{eqnarray}
and apply the relation \cite{DD84}:
\begin{equation}
 S_2(j) = - \frac{S_1(j)}{j} + \zeta(2)
        + \int_0^1 dx \, x^{j-1} \ln x\ \ln(1-x).
\end{equation}
After some manipulation one finds:
\begin{eqnarray}
\widetilde{S}_{1,2}(n)
&=& - S_{1,2}(n) + \frac12 \zeta(2) \left( S_1(n)
   - \widetilde{S}_1(n) \right)  \nonumber\\
&+& \frac14 \int_0^1 dx\,
    \frac{x^{n/2}-1}{x(x-1)} \ln x \ln(1-x) \nonumber\\
&+& \frac12 \int_0^1 dx\,
    \frac{x^{n/2}-1}{x-1} \mbox{Li}_2(x)    \nonumber\\
&+& \frac18 \int_0^1 dx\,
    \frac{2-2x^{n/2}-nx^{n/2}+nxx^{n/2}}{(x-1)^2}
    \ln x\ \mbox{Li}_2(x)  \nonumber\\
&+& \frac14 \int_0^1 dx\,
    x^{-3/2} \left(\frac12 \ln x -1\right)
    \ln(1-x) \ln\left(
    \frac{1-\sqrt{x}}{1+\sqrt{x}} \right)  \nonumber\\
&+& \frac14 \int_0^1 dx\,
    \frac{x^{n/2-1}}{n+1} \left(\frac12 \ln x -1
        \right) \ln(1-x)  \nonumber\\
& & \mbox{\ \ \ \ \ \ } \times \left[ _2F_1(1,n+1,n+2,\sqrt{x})
        + \ \!\!_2F_1(1,n+1,n+2,-\sqrt{x}) \right],
\end{eqnarray}
where $_2F_1(a,b,c,z)$ is the hypergeometric function.
For our application it is convenient to use the integral
representation for $_2F_1$ \cite{AS68}:
\begin{equation}
_2F_1(a,b,c,z) = \frac{\Gamma(c)}{\Gamma(b)\Gamma(c-b)}
     \int_0^1 dt\, t^{b-1}(1-t)^{c-b-1}(1-tz)^{-a}.
\end{equation}
Finally, making the substitution $ t \rightarrow e^u$
one arrives at:
\begin{equation}
  _2F_1(1,n+1,n+2,z) + \ \!\! _2F_1(1,n+1,n+2,-z)
= 2(n+1)\int_{-\infty}^0
  du \frac{e^{(\Re e\, n+1)\,u}}{1 - e^{2u}z^2}
     e^{i (\Im m\, n)\, u}.
\end{equation}

\section{Factorization Scheme Dependence}
\label{fact}

In this Appendix we summarize the connection between moments of 
structure functions and parton distributions in different factorization
schemes, which in NLO is not unique.
For the moment, ${\cal F}_{2,n}^{NS}(Q^2)$, of the non-singlet part of 
$F_2(x,Q^2)$ we have:
\begin{equation}
{\cal F}_{2,n}^{NS}(Q^2)
= C_{2,n}^{NS}(Q^2)\ A_n^{NS}(Q^2), 
\end{equation}
where $C_{2,n}^{NS}(Q^2)$ is the Wilson coefficient, and $A_n^{NS}(Q^2)$ is obtained from the matrix element of the 
local operator taken between nucleon states.  
Both quantities on the r.h.s. are factorization scheme dependent. 
In the $\overline{\mbox{MS}}$ factorization scheme, for example, 
the moments of parton distributions are related to $A_n^{NS}(Q^2)$ 
simply by:
\begin{equation}
{\cal Q}_n^{NS}(Q^2) = A_n^{NS}(Q^2).
\end{equation}

In the other widely used factorization scheme, namely the DIS scheme, 
the Wilson coefficient $C_{2,n}^{NS}(Q^2)$ is absorbed into the 
definition of the moments of parton distributions themselves:
\begin{equation}
 C_{2,n}^{NS}(Q^2) \equiv 1. 
\end{equation}
Therefore ${\cal F}_{2,n}^{NS}(Q^2)$ in the DIS scheme is defined to 
have the same form as in the simple parton model with free quarks.

In analogy, the singlet part of the structure function $F_2(x,Q^2)$ 
in the DIS scheme must also coincide with the simple parton model 
result \cite{AE78}:
\begin{equation}
{\cal F}^{S}_{2,n}(Q^2) = \Sigma_n^{{\rm DIS}}(Q^2).
\end{equation}
Here the Wilson coefficient, as well as the gluonic contribution, 
are absorbed into the singlet quark distribution.  
However, this procedure does not uniquely fix the gluon distribution.  
For this one enforces momentum conservation, which leads to the 
relation \cite{AE78}:
\begin{equation}
\Sigma_{n=2}^{{\rm DIS}}(Q^2)
+ {\cal G}_{n=2}^{{\rm DIS}}(Q^2)
= \Sigma_{n=2}^{\overline{{\rm MS}}}(Q^2)
+ {\cal G}_{n=2}^{\overline{{\rm MS}}}(Q^2).
\end{equation}
To fully define the moments of the gluon distribution one extends 
this equality to all $n$.

Since the anomalous dimensions are in a one-to-one correspondence 
with the renormalization constants of local operators, 
the evolution equations are affected by transformations 
to other factorization schemes.  
Because the renormalization constants depend on the renormalization 
scheme chosen to renormalize the local operators ($\equiv$ 
factorization scheme), the anomalous dimensions are factorization 
scheme dependent.
Transforming from the $\overline{\mbox{MS}}$ to the DIS scheme, 
the anomalous dimensions change according to \cite{Di88,GR82}:
\begin{eqnarray}
 \mbox{\boldmath$\gamma$}_n^{(1),{\rm DIS}}
&=& \mbox{\boldmath$\gamma$}_n^{(1),
 \overline{{\rm MS}}} + 2\beta_0 {\bf Z}_n^{(1)}
 -[\mbox{\boldmath$\gamma$}_n^{(0)},{\bf Z}_n^{1}], \\
{\bf Z}_n^{(1)}
&=& \left\{ \begin{array}{cl}
              C_{2,n}^{(1),q} & \mbox{nonsinglet},   \nonumber\\
              \left( \begin{array}{rr}
              C_{2,n}^{(1),q} & C_{2,n}^{(1),G}\\
              - C_{2,n}^{(1),q} & - C_{2,n}^{(1),G} \end{array} \right)
              & \mbox{singlet}.
            \end{array}
    \right.
\label{trafo}
\end{eqnarray}
{}From this follow the $Q^2$ evolution equations in the DIS scheme.
%


\newpage

\begin{figure}
\centering{\ \psfig{figure=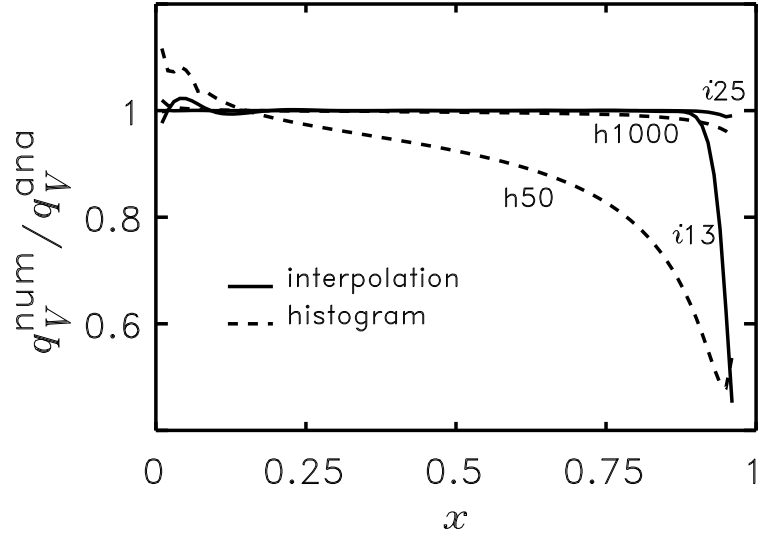,height=8.5cm}}
 \caption{Ratio of evolution results for $q_V = u_V + d_V$ 
         obtained by calculating moments numerically 
         ($q_V^{\rm num}$) and analytically ($q_V^{\rm ana}$).
         For comparison, two methods for calculating the moments 
         numerically (interpolation and histogram) have been 
         applied, using various numbers of points in Bjorken-$x$ 
         (given on top of the lines accompanied by a letter 
         indicating the method used).}
\label{F1}
\end{figure}

\newpage

\begin{figure}
\centering{\ \psfig{figure=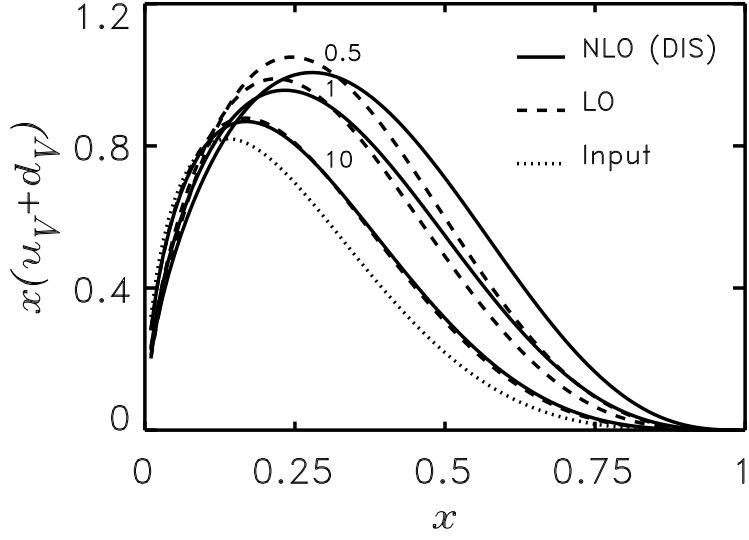,height=8.5cm}}
 \caption{Comparison of LO and NLO (DIS-scheme) evolution of
          the total valence distribution $x(u_V+d_V)$. 
          Input taken from the latest CTEQ fit \protect\cite{CTEQ} 
          is evolved from $Q^2=100$ GeV$^2$ down to different values 
          of $Q^2$ 
          (given on top of the corresponding lines).}
\label{F2}
\end{figure}

\begin{figure}
\centering{\ \psfig{figure=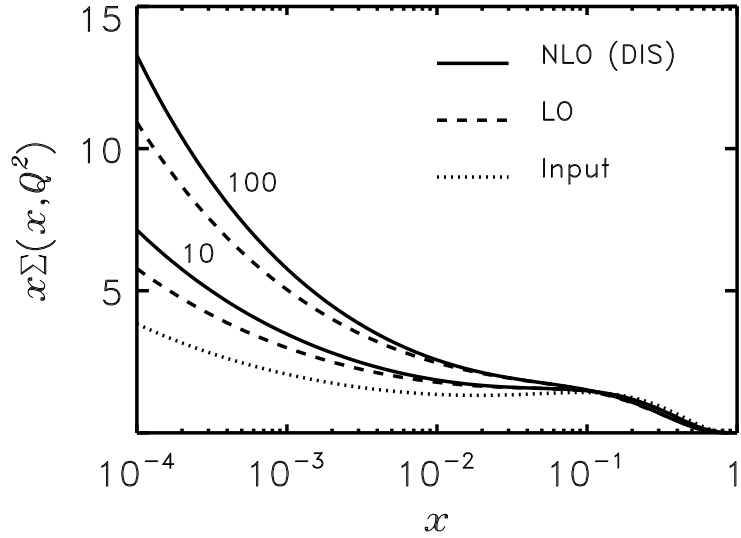,height=8.5cm}}
 \caption{Comparison of LO and NLO (DIS scheme) evolution of
          the singlet distribution $x\Sigma(x,Q^2)$.
          Values of $Q^2$ are attached to corresponding 
          pairs of lines.}
\label{F3}
\end{figure}

\begin{figure}
\centering{\ \psfig{figure=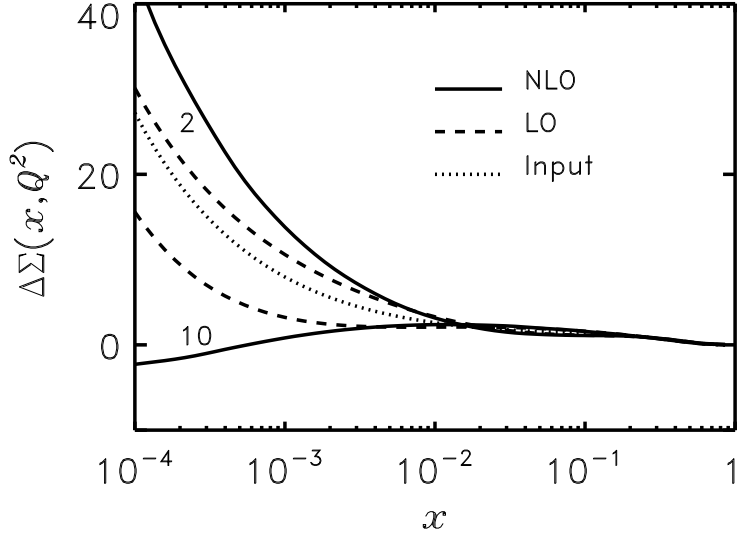,height=8.5cm}}
 \caption{Comparison of LO and NLO evolution of the polarized
          singlet distribution $x\Delta\Sigma(x,Q^2)$.
          Input is taken from Ref.\protect\cite{GS95} (Set A).
          Values of $Q^2$ are given for each pair of lines.}
\label{F4}
\end{figure}

\begin{figure}
\centering{\ \psfig{figure=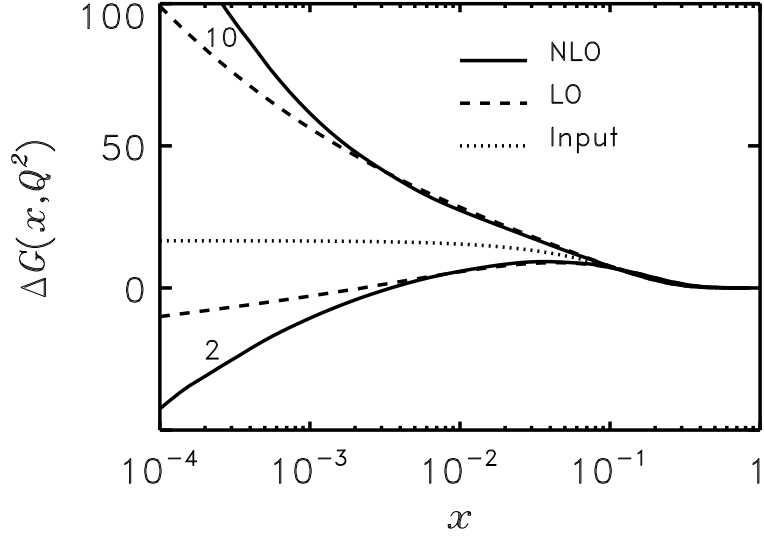,height=8.5cm}}
 \caption{Comparison of LO and NLO evolution of the polarized
          gluon distribution $x\Delta G(x,Q^2)$.
          Values of $Q^2$ are attached to corresponding 
          pairs of lines.}
\label{F5}
\end{figure}

\begin{figure}
\centering{\ \psfig{figure=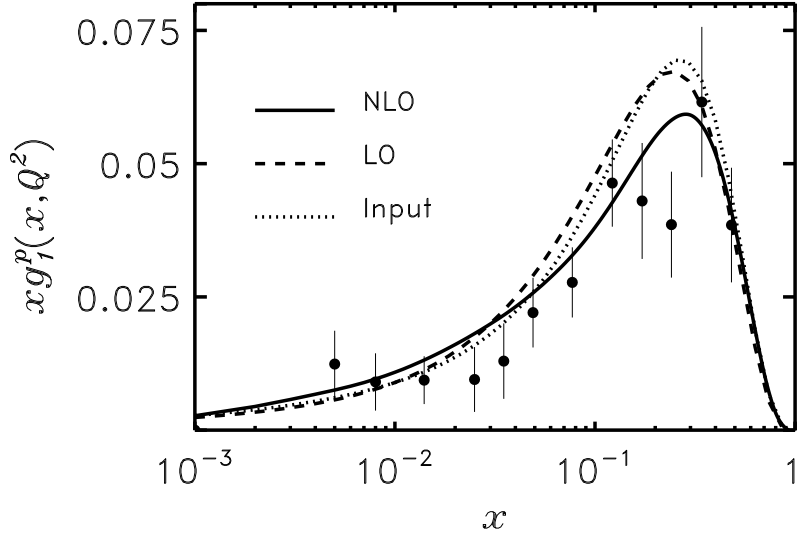,height=8.5cm}}
 \caption{NLO effects on the shape of the structure function 
          $g_1^p$. Data shown are from the SMC \protect\cite{SMCP}
          at an averaged value of $\langle Q^2 \rangle=10$ GeV$^2$,
          to which also the input has been evolved.}
\label{F6}
\end{figure}

\begin{figure}
\centering{\ \psfig{figure=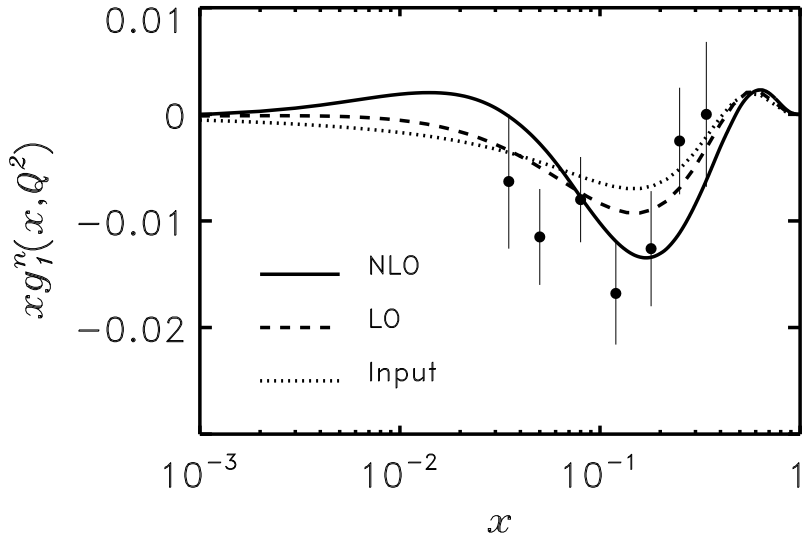,height=8.5cm}}
 \caption{NLO effects on the shape of the structure function 
          $g_1^n$. Input is evolved to 2 GeV$^2$ for comparison 
          with the SLAC E142 data \protect\cite{E142}.} 
\label{F7}
\end{figure}

\end{document}